\newcounter{myctr}
\def\myitem{\refstepcounter{myctr}\bibfont\noindent\ifnum\themyctr>9\else\phantom{0}\fi\hangindent17pt\themyctr.\enskip}

\documentclass{ws-ijqi}
\usepackage{graphicx}
\begin{document}

\markboth{Chitra Shukla et al.}
{Secure Quantum Communication with Orthogonal States }

\catchline{}{}{}{}{}

\title{Secure Quantum Communication with Orthogonal States  }

\author{Chitra Shukla}
\address{Jaypee Institute of Information Technology,
A-10, Sector-62, Noida, India\\
chitrashukla07@gmail.com}

\author{Anindita Banerjee}

\address{Bose Institute,Centre of Astro particle Physics and Space Science, EN Block, Sector - V, Salt Lake,
Kolkata, India\\
aninditabanerjee.physics@gmail.com}

\author{Anirban Pathak \footnote{Corresponding author. This article is a modified version of the tutorial lecture delivered by Anirban Pathak at International
Program on Quantum Information (IPQI-2014), February 17-28, 2014, Institute of Physics, Bhubaneswar, India.}}
\address{Jaypee Institute of Information Technology,
A-10, Sector-62, Noida, India\\
anirban.pathak@jiit.ac.in}

\author{R. Srikanth}
\address{Poornaprajna Institute of Scientific Research, Sadashivnagar, Bengaluru- 560080, India \\
srik@poornaprajna.org}
\maketitle

\begin{history}
\received{Day Month Year}
\end{history}

\begin{abstract}
In majority of protocols of secure quantum communication (such as,
BB84, B92, etc.), the unconditional security of the protocols
are obtained by using conjugate coding (two or more mutually unbiased bases). Initially
all the conjugate-coding-based protocols of secure quantum communication
were restricted to quantum key distribution (QKD), but later on they
were extended to other cryptographic tasks (such as, secure direct
quantum communication and quantum key agreement). In contrast to the
conjugate-coding-based protocols, a few completely orthogonal-state-based
protocols of unconditionally secure QKD (such as, Goldenberg-Vaidman
(GV) and N09) were also proposed. However, till the recent past orthogonal-state-based
protocols were only a theoretical concept and were limited to QKD.
Only recently, orthogonal-state-based protocols of QKD are experimentally
realized and extended to cryptographic tasks beyond QKD. This paper
aims to briefly review the orthogonal-state-based protocols of secure
quantum communication that are recently introduced by our group and
other researchers.
\end{abstract}


\section{Introduction}

Quantum cryptography is now 30 years old as it was first introduced in 1984 when Bennett and Brassard \cite{bb84}
proposed the first protocol of quantum key distribution (QKD) which
is now known as BB84 protocol. This pioneering work drew considerable
attention of the entire cryptography community as it was successful
in achieving unconditional security, a much desirable feat that is
never achievable in the classical cryptography. To be precise, all
the classical cryptographic protocols including the widely used RSA
protocol are secure only under some assumptions, whereas quantum cryptographic
protocols are unconditionally secure. Due to this existing feature
of QKD, Bennett and Brassard's initial proposal was followed by a
large number of alternate protocols of QKD \cite{ekert,b92,vaidman-goldenberg}.
The applicability of early protocols of quantum cryptography \cite{bb84,ekert,b92,vaidman-goldenberg}
were limited to QKD. However, it was soon realized that quantum states
can be employed for other cryptographic tasks, too. For example, quantum
states can be used for quantum secret sharing (QSS) of classical secrets
\cite{Hillery}\textit{, }\textit{\emph{deterministic secure quantum
communication}} (DSQC) \cite{dsqc_summation,dsqcqithout-max-entanglement,dsqcwithteleporta,entanglement swapping,hwang-hwang-Tsai,reordering1,cao and song,the:high-capacity-wstate},
quantum secure direct communication (QSDC) \cite{Long and Liu,ping-pong,for PP,lm05},
quantum dialogue \cite{ba-an,qd}, quantum key agreement (Ref. \refcite{qka-chitra}
and references therein), etc. Reviews on these topics, present challenges
and future prospects of secure quantum communication can be found
in Refs. \cite{Gisin's review,review,my book,Marco-ground breaking}.

The unconditional security of the existing protocols are usually claimed
to be obtained using different approaches and different quantum resources like, single
particle states \cite{bb84,b92,ping-pong,lm05,CL}, entangled state
\cite{ekert}, teleportation \cite{dsqcwithteleporta}, entanglement
swapping \cite{entanglement  swapping}, rearrangement of order of
particles \cite{reordering1,the:C.-W.-Tsai}, etc. Although these
protocols differ with each other with respect to the procedure followed
and the quantum resources used, the security of all these protocols
of secure quantum communication essentially arises from the use of
conjugate coding (i.e., from the quantum non-commutativity or equivalently
from the use of two or more mutually unbiased bases (MUBs)) as in
all these protocols the existence of an eavesdropper is traced by
measuring verification qubits in 2 or more MUBs. Thus all these protocols may be viewed as conjugate-coding-based
protocols of quantum communication, alternatively these protocols
may be referred to as BB84-type protocols of quantum communication.

The existence of such a large number of conjugate-coding-based protocols
of quantum communication leads to a fundamental question: Is conjugate
coding essential for unconditionally secure quantum communication?
The answer is ``no''. Specifically, it is possible to design protocols
of secure quantum communication using orthogonal states alone. Thus
we can design protocols of secure quantum communication using orthogonal
states for encoding of information, decoding of information and eavesdropping
check i.e., using a single basis for implementation of the entire
protocol without involving any use of two or more MUBs or conjugate
coding. First such orthogonal-state-based protocol was reported in
1995 by Goldenberg and Vaidman \cite{vaidman-goldenberg} and subsequently
a few other orthogonal-state-based protocols of QKD were reported
\cite{N09,guo-shi,koashi-imoto}. However, till recent past activities on orthogonal-state
based protocols of quantum communication were limited to QKD and theoretical
studies alone. Only recently a set of exciting experiments on orthogonal-state-based
protocols of quantum communication have been reported \cite{GV-experiment,N09-expt-1,N09 expt-2,N09 expt-3}.
Further, new orthogonal-state-based protocols are proposed for quantum
cryptographic tasks beyond QKD \cite{qka-chitra,beyond-gv,preeti-arxiv,dsqc-ent swap,salih,cf-more efficient  n09,cfl-info-trans,cf-srikanth-cert-authentication,cf-Salih- trans-of-qubit,cfl-ent-distribution,cf-database-query,cf-salih-tripartite-qcrypt}.
These orthogonal-state-based proposals can be broadly classified in
two classes: (i) GV-type protocols which are analogous to the original
GV-protocol and in which transmission of qubits that carry secret
information through the quantum channel is allowed, but the information
is protected from the eavesdropping by geographically separating an
orthogonal state into two or more quantum pieces that are not simultaneously
accessible to Eve and (ii) N09-type protocols or counterfactual protocols
that use interaction free measurement and circumvents
the transmission of information carrying qubits through the quantum
channel. GV-type protocols are mostly investigated by the present
authors and their collaborators \cite{qka-chitra,my book,beyond-gv,preeti-arxiv,dsqc-ent swap}.
Specifically, we have shown that it is feasible to construct orthogonal-state-based
protocols of QKA \cite{qka-chitra}, QSDC and DSQC \cite{beyond-gv,preeti-arxiv,dsqc-ent swap}.
Practically, we have established that all the secure quantum communication
tasks that can be performed using two or more MUBs can also be achieved
by using single basis. Similarly, much progress has recently been
made in designing of counterfactual (i.e., N09-type) protocols. For
example, in 2013, Salih et al. have claimed to design a counterfactual
protocol of  direct quantum  communication \cite{salih}. The claim was
subsequently criticized by Vaidman \cite{vaidman-on-salih} and the
criticism lead to a very interesting debate on the issue \cite{response-to-vaidman}.
Further, recently Salih has also proposed counterfactual protocols
for transportation of an unknown qubit \cite{cf-Salih- trans-of-qubit}
and tripartite quantum cryptography \cite{cf-salih-tripartite-qcrypt},
Guo et al. have proposed protocol of counterfactual entanglement distribution
\cite{cfl-ent-distribution}, Guo et al. proposed protocol of counterfactual
information transfer \cite{cfl-info-trans}, Sun and Wen have proposed
a modified N09 protocol \cite{cf-more efficient  n09} which is more
efficient than the actual N09 protocol and some of the present authors
proposed protocols of counterfactual certificate authentication \cite{cf-srikanth-cert-authentication}
and semi-counterfactual QKD \cite{Akshata1}. These exciting developments
of recent past motivated us to briefly review these recent achievements
with specific attention to works of our group. 

Here we will briefly review a set of existing orthogonal-state-based protocols and describe a trick that helps us to transform BB84-type
protocols into Goldenberg-Vaidman (GV) type \cite{vaidman-goldenberg}
protocols, which uses only orthogonal states for encoding, decoding
and error checking, as was done in the original GV protocol of QKD.
Subsequently, we will describe two orthogonal-state-based protocols
of quantum communication introduced by us and briefly describe how
they can be extended. These two orthogonal-state-based protocols are
fundamentally different from conjugate-coding-based (BB84-type) protocols
as their security does not depend on noncommutativity. Consequently,
they are very important from the foundational perspective.

The trick that can transform BB84-type protocols into GV-type protocol
requires the rearrangement of orders of particles or \textit{permutation
of particles} (PoP). As PoP plays a very crucial role in our protocol,
it would be apt to note that this technique was first introduced by Deng and Long in 2003, while they proposed a protocol of QKD based on this technique \cite{PoP}. Subsequently, a DSQC protocol based on the rearrangement
of orders of particles was proposed by Zhu et al\emph{.} \cite{reordering1}
in 2006. However, it was shown to be insecure under a Trojan-horse
attack by Li et al. \cite{dsqcqithout-max-entanglement}. In Ref.
\cite{dsqcqithout-max-entanglement}, Li et al. had also provided
an improved version of Zhu et al. protocol that is free from the above
mentioned Trojan-horse attack. Thus we may consider Li et al. protocol
as the first unconditionally secure protocol of DSQC based on PoP.
Recently, many PoP based protocols are proposed (See \cite{my book}
and references therein). Specifically, many such PoP-based protocols
of quantum communication have been proposed in recent past. For example,
Banerjee and Pathak \cite{Anindita}, Shukla, Banerjee and Pathak
\cite{with chitra-ijtp}, Yuan et al\emph{.} \cite{the:high-capacity-wstate}
and Tsai et al\emph{.} \cite{the:C.-W.-Tsai} have recently proposed
PoP-based protocols of direct secure quantum communication. In what
follows, we will see that PoP provides us a useful tool for the  generalization
of the original GV protocol into corresponding multipartite version.

The remaining part of the present paper is organized as follows, in
Section \ref{sec:A-chronological-history}, we briefly review the
development of orthogonal-state-based secure quantum communication
until now by providing a chronological history of developments of
protocols of orthogonal-state-based secure quantum communication and
their experimental verifications. In Section \ref{sec:Role-of-nocloning},
we discuss the role of no-cloning and randomness in secure communication
and with some specific examples show that it is possible to transform
all BB84-type protocols of secure quantum communication to corresponding
GV-type protocols. Finally, the paper is concluded in Section \ref{sec:Conclusions}.

\section{A chronological history of protocols of orthogonal-state-based secure
quantum communication and their experimental verification\label{sec:A-chronological-history}}
\begin{figure}[pb]
\includegraphics[scale=0.35]{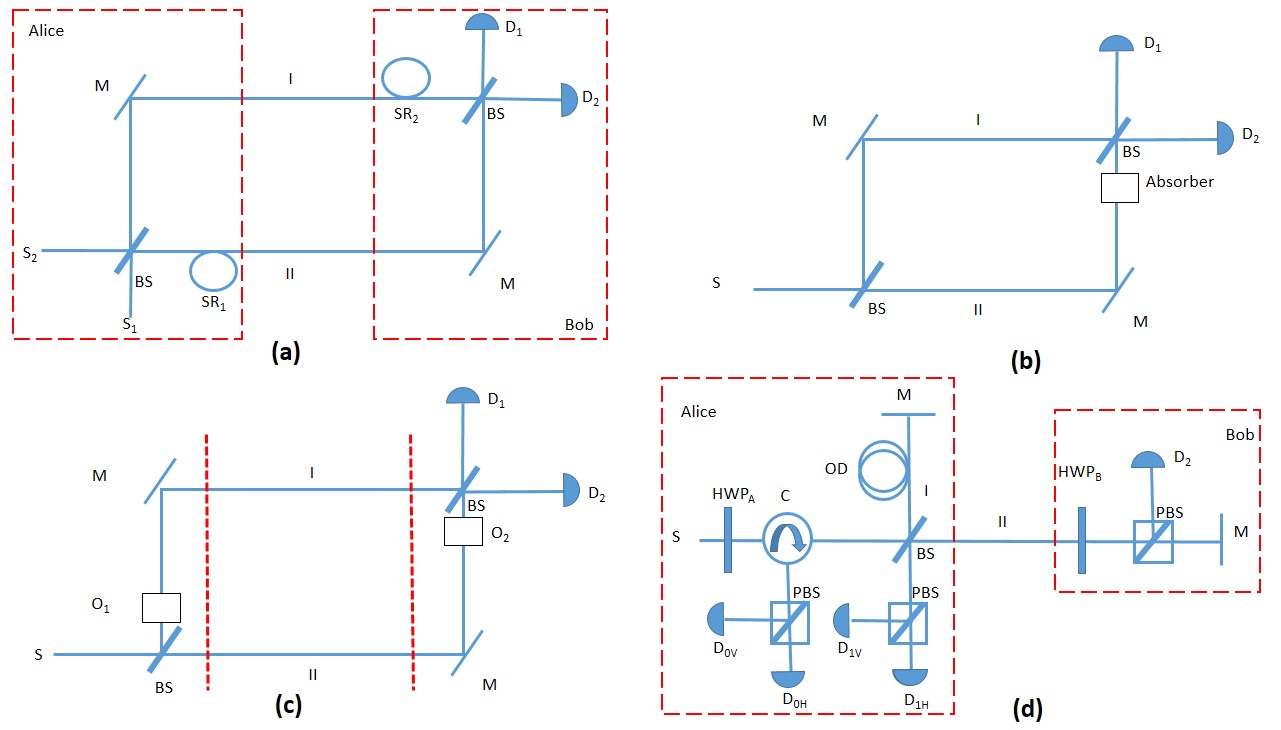}

\caption{\label{fig:1}(a) A schematic diagram of a modified Mach-Zehnder interferometer
that can be used to implement GV protocol [4] if the symmetric beam spliters used as 50:50, otherwise (i.e., if the symmetric beam splitters are not 50:50) the same device implements Koashi-Imoto protocol [29].
Here SR$_{i}$ denotes a delay. (b) A schematic diagram of a Mach-Zehnder
interferometer that can be used to implement Elitzur and Vaidman's
idea of interaction-free measurement or quantum interrogation [54].
Here the absorber is an active bomb that blasts when receives a photon.
(c) A schematic diagram of Mach-Zehnder interferometer that can be
used to realize orthogonal-state-based protocol of Guo-Shi [28]
which uses interaction-free measurement. Here ${\rm O_{1}}$ and ${\rm O_{2}}$
are the obstacles that are randomly inserted by Alice and Bob, respectively.
(d) A schematic diagram of the experimental setup used in Refs. [24. 32]
to implement N09 protocol [27] of counterfactual QKD. In all the diagrams
BS, M, C, PBS, HWP, D and OD represent beam splitter, mirror, circulator,
polarizing beam splitter, half wave plate, detector and optical delay,
respectively.}
\end{figure}

\begin{description}
\item [{1995:}] All the protocols of quantum cryptography proposed until
1995 were based on nonorthogonal states and security of those protocols
arose directly or indirectly through noncommutativity, but in 1995,
Goldenberg and Vaidman \cite{vaidman-goldenberg} proposed a completely
orthogonal-state-based protocol of QKD, where the security arises
due to duality (for single particle). This was the birth of orthogonal-state-based
protocol of quantum cryptography. Interestingly, the fact that GV
protocol is fundamentally different from the BB84-type protocol was
questioned by Peres \cite{peres criticism}. However, Goldenberg and
Vaidman successfully defended their work \cite{gv reply} and established
the fact that this orthogonal-state-based protocol is fundamentally
different from the conventional BB84-type protocol. In the next section
we have briefly described this protocol and have shown that the protocol
uses a slightly modified Mach-Zehnder interferometer (See Fig. \ref{fig:1}a
).
\item[{1997:}] Koashi and Imoto \cite{koashi-imoto} generalized the GV protocol and proposed a protocol similar to GV protocol, but does not require random sending time.
\item [{1998:}] Mor \cite{tal mor} showed that it is not always possible
to clone orthogonal states. Specifically, an orthogonal state cannot be cloned if  the full state cannot
be accessed at the same time. Using this idea, Mor provided a clear
and innovative explanation of the origin of security of GV protocol.
\item [{1999:}] Four years after the introduction of first orthogonal-state-based
QKD protocol (i.e., GV protocol), Guo and Shi \cite{guo-shi} proposed
the second orthogonal-state-based protocol of QKD using the concept
of interaction-free measurement or quantum interrogation \cite{Bomb testing},
an idea that was introduced earlier by Elitzur and Vaidman in context
of a very interesting hypothetical situation in which some of the
active bombs can be separated from the inactive ones without directly
observing the active bombs (i.e., without sending any photon to the
isolated active bombs which blasts when receives a photon, whereas
inactive bombs does not show any response on receiving a photon).
Actually, the bombs are placed in the lower arm of a Mach-Zehnder
interferometer and a single photon is sent through the input port
(See Fig. \ref{fig:1}b). With 50\% probability the single photon
travels through the upper arm of the interferometer. Even in these
50\% cases, if we have an active bomb (thus a detector) in the lower
arm, the interference is destroyed as we obtain the which path information,
and consequently in half of these incidents (i.e., 25\% of the total)
the detector present at the output port of the interferometer that
does not click in absence of any detector in the lower arm would
click. As a consequence we will be able to detect 25\% of the active
bombs without blasting them. Thus, in brief, the presence of the obstacle
(active bomb) disrupts the destructive interference that would otherwise
occur and thereby reveal its presence. Guo and Shi modified the idea
and in their protocol Alice (Bob) randomly inserts an absorber in
upper (lower) arm of the interferometer (See Fig. \ref{fig:1}c).
Form the clicks of the upper detector which does not click in absence
of the detector, Bob knows that one of the absorber was present in
one of the arm. In these cases he discloses that his upper detector
has been clicked. As Alice (Bob) knows whether she (he) has inserted
the absorber, using the observation of Bob she (he) can conclude whether
Bob (Alice) has inserted the absorber or not and subsequently use
this to form a key using a pre-decided rule: presence of Alice's (Bob's)
absorber implies bit value $0\,(1).$ Anyway, Guo and Shi's effort
was the first step towards orthogonal-state-based counterfactual QKD
and in recent years interaction-free measurement is frequently used
as a tool for the designing of counterfactual quantum cryptographic
protocols.
\item [{2009:}] A protocol of counterfactual QKD (orthogonal-state-based)
was proposed by Noh in 2009 \cite{N09} using the Elitzur and Vaidman's
idea of interaction-free measurement. This protocol of QKD is now
known as N09 protocol or counterfactual protocol. This protocol led
to many subsequent counterfactual protocols of secure quantum communication.
The beauty of this protocol and other counterfactual protocols is
that a secure key is distributed (or other cryptographic task is achieved)
without transmitting a particle that carries secret information through
the quantum channel. Interestingly, in GV, Koashi-Imoto and Guo-Shi protocols Mach-Zehnder
interferometer was used, but in this protocol a Michelson interferometer
is used (See Fig. \ref{fig:1}d)%
\footnote{\begin{description}
\item [{\textmd{Interested}}] readers may refer to Refs. \cite{Marco-ground breaking,GV-experiment}
for detail description of the setup.\end{description}
}.
\item [{2010:}] Sun and Wen \cite{cf-more efficient  n09} improved the
original N09 protocol by providing analogous counterfactual protocol
with higher frequency. In the same year, Avella et al. experimentally
implemented GV protocol \cite{GV-experiment}. To the best of our
knowledge this was the first ever experimental demonstration of orthogonal-state-based
protocol of QKD.
\item [{2011:}] Experimental realization of N09 protocol was reported shortly
after realization of GV protocol. Precisely, in 2011, Ren et al. reported
experimental realization of N09 protocol \cite{N09-expt-1}.
\item [{2012:}] Soon after Ren et al.'s work two more groups reported experimental
realization of N09 protocol. Specifically, Brida et al. \cite{N09 expt-2}
and Liu et al. \cite{N09 expt-3}, independently implemented this
protocol of counterfactual quantum communication. In theoretical front,
some of the present authors generalized the single particle GV protocol
to multipartite case \cite{preeti-arxiv} and showed that GV-type
protocol can be used for secure direct communication and established
that while in GV the encoding states are perfectly indistinguishable,
in the bi-partite case, they are partially distinguishable, leading
to a qualitatively different kind of information-vs-disturbance trade-off
and also options for Eve in the two cases. Further, generalizing the
idea we had also established that GV-type protocol of DSQC, QSDC and
QKD can be realized using arbitrary quantum states \cite{beyond-gv}.
Essential ideas that lead to these multipartite GV-type protocols
will be explained briefly in the next section.
\item [{2013:}] While the above counterfactual protocols are probabilistic,
H. Salih et al. proposed a protocol for counterfactual  direct 
quantum communication \cite{salih}. This work of Salih et al., led
to an interesting debate and whether the protocol is counterfactual
for only one of the two bit values has been controversial \cite{vaidman-on-salih}.
This protocol efficiently uses chained quantum Zeno effect and an
arrangement of sequence of Mach-Zehnder interferometers, where each
of the Mach-Zehnder interferometer essentially uses Elitzur and Vaidman
setup for interaction free measurement. In the same year, Zhang et
al. proposed a counterfactual protocol of private database queries
\cite{cf-database-query} which also uses a similar setup of sequence
of Mach-Zehnder interferometer. Further, the applicability of GV-type
orthogonal-state-based protocols of secure quantum communication was
extended by some of the present authors to quantum key agreement (QKA)
where Alice and Bob contribute equally to the final shared key and
none of them can control the final key \cite{qka-chitra}.
\item [{2014:}] 2014 is the most active year in the history of orthogonal-state-based
secure quantum communication. In this year many interesting results
appeared. Here we list a few of them: (i) Guo et al. proposed a protocol
of counterfactual quantum-information transfer \cite{cfl-info-trans},
(ii) Guo et al. proposed a counterfactual protocol of entanglement
distribution \cite{cfl-ent-distribution}, (iii) Salih proposed a
multiparty (tripartite) scheme of counterfactual quantum communication
\cite{cf-salih-tripartite-qcrypt} and (iv) some of the present authors
proposed a scheme for counterfactual quantum certificate authorization
\cite{cf-srikanth-cert-authentication}.
\end{description}
In the above chronological review we have seen that majority of the
interesting developments in orthogonal-state-based secure quantum
communication happened in last few years. The development is expected
to continue and it is expected to play important role in practical
realization of secure quantum communication and also in our understanding
of quantum mechanics in general and origin of security in quantum
mechanics and post-quantum theories in particular. Keeping these facts
in mind, in the next section we briefly review role of no-cloning
theorem in realization of orthogonal-state-based protocols and also
briefly describe a few orthogonal-state-based GV-type protocols of
secure quantum communication.

\section{Role of no-cloning and randomness in secure communication and how
to transform BB84-type protocols to GV-type protocols\label{sec:Role-of-nocloning}}

It is well known that unknown quantum states cannot be cloned and
several proofs of no-cloning theorem are provided using unitary evolution
\cite{tal mor}, no-signaling \cite{NoCloning1}, linearity \cite{Nocloning3}.
A closer look into these proofs reveals that there exist fine differences
among these proofs and those differences lead to a fundamental question:
What nonclassical resources are required for the existence of no-cloning
theorem in a theory $T$. Recently, we have shown that no-cloning theorem
should hold in any theory possessing uncertainty and disturbance on
measurement \cite{post-quantum}. Thus we can construct post-quantum
theories with no-cloning. Without going into detail of those theories,
let us try to follow a simpler argument that can give us a general
perception of no-cloning theorem. To begin with let us try to address
another simple question: What distinguishes a completely stochastic
classical theory from the quantum mechanics? Clearly, in a completely
stochastic classical theory the outcomes of measurement are always
probabilistic whereas in quantum mechanics we can have a deterministic
outcome if the state to be measured is part of the basis set used
for the measurement. For example, if we measure $|0\rangle$ in $\{|0\rangle,|1\rangle\}$
basis we will always get $|0\rangle$ (thus the outcome is deterministic
as the state is part of the basis), but if we measure $|0\rangle$
in $\{|+\rangle,|-\rangle\}$ basis we will have probabilistic outcome.
We may say that the $\{|0\rangle,|1\rangle\}$ basis is \emph{special
basis} as it leads to deterministic outcome. We may now generalize
the idea and say that for measurement of a state a particular basis
set will be referred to as special basis if the state can be perfectly
measured in that basis. It is easy to recognize that existence of
special basis implies perfect measurement and thus complete information
of the state being measured. This information implies that the state
is known and thus can be cloned. In contrast, absence of special basis
implies no-cloning. As the elements of any basis set are orthogonal
to each other, two nonorthogonal states cannot be part of the same
basis set and thus cannot be cloned. However, this viewpoint does
not demand that the orthogonal states can always be cloned. Specifically,
by using geographical separation among the components of a superposition
state we can make it non-clonable. In a completely different language
this viewpoint was elaborated by Mor \cite{tal mor} in 1998. Of course
Mor's work appeared after the GV protocol, but it helped us to understand
and generalize GV protocol. Let us elaborate this point by briefly
describing GV protocol.

\subsection{Goldenberg-Vaidman (GV) protocol}

Let us consider two orthogonal states
\begin{equation}
\begin{array}{lcl}
|\psi_{0}\rangle & = & \frac{1}{\sqrt{2}}\left(|a\rangle+|b\rangle\right)\end{array}\label{eq:one-1}
\end{equation}
and
\begin{equation}
\begin{array}{lcl}
|\psi_{1}\rangle & = & \frac{1}{\sqrt{2}}\left(|a\rangle-|b\rangle\right),\end{array}\label{eq:two}
\end{equation}
where $|a\rangle$ and  $|b\rangle$ are two localized wave packets.
Further, $|\psi_{0}\rangle$ and $|\psi_{1}\rangle$ represent bit
values $0$ and $1$, respectively. Alice sends wave packets  $|a\rangle$
and  $|b\rangle$ to Bob by using two different arms of a Mach-Zehnder
interferometer as shown in the Fig. \ref{fig:1} a. Alice sends Bob
either  $|\psi_{0}\rangle$ or  $|\psi_{1}\rangle,$ but $|a\rangle$
is always sent first and $|b\rangle$ is delayed by time $\tau$.
Here traveling time ($\theta$) of wave packets from Alice to Bob
is shorter than $\tau$ . Thus $|b\rangle$ enters the communication
channel only after $|a\rangle$ is received by Bob. Consequently,
both the wave packets $|a\rangle$ and $|b\rangle$ (i.e., the entire
superposition) are never found simultaneously in the transmission
channel. This geographic separation between $|a\rangle$ and $|b\rangle$
restricts Eve from measuring the state communicated by Alice in $\left\{ |\psi_{0}\rangle,|\psi_{1}\rangle\right\} $
basis. In fact, this geographic separation method compels Eve to measure
the state communicated by Alice either in $\left\{ |a\rangle,|b\rangle\right\} $
basis or in some suitably constructed positive-operator valued measure (POVM). Thus the geographic separation
ensures unavailability of special basis and thus implies no-cloning
and security of GV protocol. This is how one can look at the security
of GV protocol using the concept of special basis or the idea of Mor
\cite{tal mor}. Although the special basis is not available to Eve,
it is available to Bob as Bob delays $|a\rangle$ by $\tau$ and recreates
the superposition state sent by Alice after he receives $|b\rangle$
(cf. Fig. \ref{fig:1} a). In order to restrict Eve to perform the
similar operation (i.e., to delay $|a\rangle$ till the arrival of
$|b\rangle)$ Alice and Bob need to perform following tests:
\begin{enumerate}
\item Alice and Bob compare the receiving time  $t_{r}$ with the sending
time  $t_{s}$ for each state to ensure that Eve cannot delay  $|a\rangle$
and wait for  $|b\rangle$ to reach her so that she can do a measurement
in $\left\{ |\psi_{0}\rangle,|\psi_{1}\rangle\right\} $. Specifically,
Alice and Bob checks that $t_{r}=t_{s}+\theta+\tau.$ \\
This test ensures that Eve cannot delay a wave packet, but it does
not stop her from replacing a wave packet by a fake wave packet.
The following test detects such an attack.
\item Alice and Bob look for changes in the data by comparing a portion
of the transmitted bits with the same portion of the received bits.
\end{enumerate}
It is important to note that sending time in GV protocol must be random.
Otherwise, Eve can prepare a fake state in $|\psi_{0}\rangle$ and
send the fake  $|a\rangle$ to Bob at the known arrival time. Eve
can keep the original $|a\rangle$ and the fake $|b\rangle$ wave
packets with her till the arrival of original $|b\rangle$. When $|b\rangle$
arrives then she measures the original state. If the measurement yields
$|\psi_{0}\rangle$ then she sends the fake wave packet $|b\rangle$
to Bob. Otherwise, she corrects the phase of the fake wave packet
and sends $-|b\rangle$ to Bob. If we assume that the time required
for Eve's measurement is negligible, then following this procedure
Eve can obtain the key without being detected. Interestingly, this requirement of random sending time can be circumvented just by replacing the 50:50 beam spliters present in the GV setup (cf. Fig.  \ref{fig:1} a ) by identical beam splitters having $R\neq T$, where
$R$ and $T$ are reflectivity and transmissivity, respectively. This small change in GV setup ( \ref{fig:1} a) turns it into Koashi-Imoto \cite{koashi-imoto} protocol. 

In the above we have already seen that it is possible to separate
two pieces of orthogonal state and that leads to unavailability of
special basis and thus no-cloning and orthogonal-state-based QKD.
In what follows we will show that validity of GV-type protocol is
not limited to single particle case and QKD, it can be easily generalized
to multipartite case and to design protocols of DSQC and QSDC. Before
we describe an orthogonal-state-based protocol of secure direct quantum
communication, we wish to note that GV in its original form is a protocol
of QKD only and it cannot be directly used for secure direct quantum
communication. Keeping this in mind, let us first describe a conjugate
coding based protocol of secure direct quantum communication. The
protocol is popularly known as ping-pong (PP) protocol \cite{ping-pong}
and is described in the following section.

\subsection{Ping-pong and modified ping-pong protocols}

Ping-pong (PP) protocol which was introduced by ${\rm Bostr\ddot{o}m}$
and Felbinger in 2002 \cite{ping-pong} is a protocol of QSDC and
it may be described briefly as follows \cite{my book}:
\begin{description}
\item [{PP1}] Bob prepares $n$ copies of the Bell state $|\psi^{+}\rangle\equiv\frac{1}{\sqrt{2}}(|00\rangle+|11\rangle)_{AB}$
(i.e., $|\psi^{+}\rangle^{\otimes n}$), and transmits all the first
qubits of the Bell pairs to Alice, keeping all the second particles
with himself.
\item [{PP2}] Alice randomly selects a set of $\frac{n}{2}$ qubits from
the string received by her as a verification string, and applies the
BB84 subroutine%
\footnote{BB84 subroutine means eavesdropping is checked by following a procedure
similar to that adopted in the original BB84 protocol. Specifically,
BB84 subroutine implies that Alice (Bob) randomly selects half of
the qubits received by her (him) to form a verification string. She
(He) measures verification qubits randomly in $\left\{ 0,1\right\} $
or $\left\{ +,-\right\} $ basis and announces the measurement outcome,
position of that qubit in the string and the basis used for the particular
measurement. Bob (Alice) also measures the corresponding qubit using
the same basis (if needed) and compares his (her) results with the
announced result of Alice (Bob) to detect eavesdropping.%
} on the verification string to detect eavesdropping. If sufficiently
few errors are found, they proceed to the next step; else, they return
to  the previous step.
\item [{PP3}] Alice randomly selects half of the unmeasured qubits as verification
string for the return path and encodes her message in the remaining
$\frac{n}{4}$ qubits using following rule: Alice does nothing to
encode 0 on a message qubit, and applies an $X$ gate to encode 1.
After completion of the encoding operation, she sends all the $\frac{n}{2}$
qubits of her possession to Bob.
\item [{PP4}] Alice discloses the coordinates of the verification qubits
after receiving authenticated acknowledgment of receipt of all the
qubits from Bob. Bob applies the BB84 subroutine on the verification
qubits and computes the error rate. If sufficiently few errors are
found, they proceed to the next step; else, they return to  PP1.
\item [{PP5}] Bob performs Bell-state measurements on the remaining Bell
pairs, and decodes the message.
\end{description}
If in \textbf{PP3} Alice has encoded 0 then Bob will obtain $|\psi^{+}\rangle$
(the same as he had sent) in \textbf{PP5}, otherwise he will receive
$|\phi^{+}\rangle.$ Since $|\psi^{+}\rangle$ and $|\phi^{+}\rangle$
are orthogonal a Bell measurement will deterministically distinguish
$|\psi^{+}\rangle$ and $|\phi^{+}\rangle$ and consequently decode
the message encrypted by Alice. This two-way protocol is referred
to as the ping-pong protocol as the travel qubit moves from Bob to
Alice and comes back just like a table tennis (ping-pong) ball which
moves back and forth between two sides of the table. It is easy to
observe that in the original PP protocol full power of dense coding
is not used. Alice could have used $I,\, X,\, iY$ and $Z$ to encode
$00,01,10$ and $11$ respectively and that would have increased the
efficiency of ping-pong protocol. This is so because the same amount
of communication would have successfully carried two bits of classical
information. This fact was first formally included in a modified PP
protocol proposed by Cai and Li in 2004 \cite{CL}. In fact, in principle
any entangled state can be used to design a ping-pong type protocol
for QSDC.

Here it is interesting to observe that in the above version of PP
protocol (and in CL protocol) encoding and decoding of information
is done by using orthogonal states alone. However, the eavesdropping
checking is done with the help of BB84 subroutine. Thus to convert
PP protocol into an orthogonal-state-based protocol we would require
to replace BB84 subroutine by a GV-type subroutine for eavesdropping
check. While describing the role of special basis on the origin of
security of GV protocol, we have already mentioned that if we can
visualize an orthogonal state as superposition of two quantum pieces
that are geographically separable then the orthogonal state can be
transmitted in such a way that Eve can neither clone it nor measure
it without disturbing. In addition, we may note that an entangled
state is a superposition in tensor product space. Now just consider
a simple situation that Alice prepares a product of two Bell states
say $|\psi^{+}\rangle^{\otimes 2}=|\psi^{+}\psi^{+}\rangle_{1234}$ and randomly
changes the sequence of the particles and sends them to Bob over a
channel. Now Eve knows that two Bell states are sent and she has to
do a Bell measurement to know which Bell state is sent, but she does
not know which particle is entangled to which particle. Consequently,
any wrong choice of partner particles would lead to entanglement swapping
(say if Eve does Bell measurement on $13$ and/or $24$ that would
lead to entanglement swapping). Now consider that at a later time,
when Bob informs Alice that he has received 4 qubits then Alice discloses
the actual sequence of the transmitted qubits and Bob uses that data
to rearrange the qubits in his hand into the original sequence and
perform Bell measurement on them. Clearly, attempts of eavesdropping
will leave detectable traces through the entanglement swapping and
whenever Bob's Bell measurement would yield any result other than
$|\psi^{+}\rangle$ they will know there exists a Eve. Clearly, this
new eavesdropping checking subroutine is of GV-type as it uses orthogonal
states only and as it geographically separates two quantum pieces
of an orthogonal state. Further, PoP technique applied here actually
ensures that the special basis (Bell basis in this case) is not available
with Eve when the particles are in the channel, but after Alice's
disclosure of the actual sequence of the qubit, Bob obtains access
to the special basis. Once we understand the essence of this strategy,
we may generalize it to develop a GV-type subroutine as follows:
\begin{enumerate}
\item To communicate a sequence $A$ of $n$ message qubits, Alice creates
an additional sequence $D$ of $n$ decoy qubits prepared as $|\psi^{\otimes\frac{n}{2}}\rangle$.
\item She concatenates $D$ with $A$ to obtain a new sequence $P$ of
$2n$ qubits and applies a permutation operator $\Pi_{2n}$ on $P$
to yield $P^{\prime}=\Pi_{2n}P.$
\item After receiving authenticated acknowledgment from Bob that he has
received all the $2n$ qubits sent to him, Alice discloses the actual
sequence of the decoy qubits only (she does not disclose the sequence
of message qubits) so that Bob can perform Bell measurement on partner
particles (original Bell pairs) and reveal any effort of eavesdropping
through the disturbance introduced by Eve's measurements.
\end{enumerate}
As the message qubits are also randomized and as Alice does not disclose
the actual sequence till she knows that eavesdropping has not happened.
Above subroutine for eavesdropping checking which we referred to as
GV subroutine can be used to convert any BB84-type protocol of secure
quantum communication that utilizes orthogonal states for encoding
and decoding. For example, PP \cite{ping-pong}, Cai-Li \cite{CL}
and DLL \cite{DLL} protocol can be converted easily into GV-type
protocol. This idea is extensively discussed in our recent publications
\cite{qka-chitra,beyond-gv,preeti-arxiv,dsqc-ent swap}. For the completeness
of the present paper we elaborate this point here by explicitly describing
a GV-type version of PP protocol which we referred to as PP$^{{\rm GV}}$.
More detail about this protocol can be found at Refs. \cite{my book,preeti-arxiv}.

\subsection{PP$^{{\rm GV}}$ protocol}

In what follows we briefly describe the PP$^{{\rm GV}}$ protocol
introduced by Yadav, Srikanth and Pathak \cite{preeti-arxiv}. We
can convert PP protocol to PP$^{{\rm GV}}$ protocol by modifying
steps \textbf{PP1}, \textbf{PP2} and \textbf{PP4} of PP protocol described
above as follows:
\begin{description}
\item [{PP$^{{\rm GV}}$1}] Bob prepares the state $|\psi^{+}\rangle^{\otimes n}$.
He keeps half of the second qubits of the Bell pairs with himself.
On the remaining $\frac{3n}{2}$ qubits he applies a random permutation
operation $\Pi_{\frac{3n}{2}}$ and transmits them to Alice. $n$
of the transmitted qubits are Bell pairs and the remaining $\frac{n}{2}$
are the partner particles of the particles which remained with Bob.
\item [{PP$^{{\rm GV}}$2}] After receiving Alice's authenticated acknowledgment,
Bob announces $\Pi_{n}\in\Pi_{\frac{3n}{2}}$, the coordinates of
the transmitted Bell pairs. Alice measures them in the Bell basis
to determine if they are each in the state $|\psi^{+}\rangle$. If
the error detected by Alice is within the tolerable limit, they continue
to the next step. Otherwise, they discard the protocol and restart
from \textbf{PP$^{{\rm GV}}$1}.
\item [{PP$^{{\rm GV}}$4}] Alice discloses the coordinates of the verification
qubits after receiving Bob's authenticated acknowledgment of receipt
of all the qubits. Bob combines the qubits of verification string
with their partner particles already in his possession and measures
them in the Bell basis to compute the (return trip) error rate.
\end{description}
The other steps in PP remain the same.

Briefly, security in PP$^{{\rm GV}}$ and CL$^{{\rm GV}}$ arises
as follows. The reordering has the same effect as time control and
time randomization in GV. Eve is unable to apply a 2-qubit operation
on legitimate partner particles to determine the encoding in spite
of their orthogonality. Any correlation she generates by interacting
with individual particles will diminish the observed correlations
between Alice and Bob because of restrictions on shareability of quantum
correlations \cite{preeti-arxiv}. It is not our purpose to discuss
the security of the protocol in detail here. Interested readers may
found detailed discussions on the security of PP$^{{\rm GV}}$ in
Refs. \cite{beyond-gv,preeti-arxiv}. The PP$^{{\rm GV}}$ protocol
of Yadav, Srikanth and Pathak was the first ever orthogonal-state-based
protocol of QSDC.

PP protocol described above is a two way protocol in the sense that
the qubits travel in both the direction (i.e., from Alice to Bob and
Bob to Alice). However, it is possible to modify them into one-way
protocols. A very interesting one-way protocol known as DLL protocol
was introduced by Deng, Long and Liu in 2003 \cite{DLL}. This protocol
can be obtained by modifying CL protocol. In what follows we will
describe DLL protocol and subsequently modify that to a GV-type protocol
which we refer to as DLL$^{{\rm GV}}$. A relatively detailed description
of this protocol can be found at Ref. \cite{my book}.

Before we describe DLL protocol we may note that after PP1, Alice
and Bob share entanglement. To share an entanglement it is not required
to be created by Bob as in PP protocol, even Alice can create an entangled
state and send a qubit to Bob. Let us modify the first step of PP
protocol and see what happens.

\subsection{\label{sub:The-Deng-protocol} DLL protocol }
\begin{description}
\item [{DLL1}] Alice prepares the state $|\psi^{+}\rangle^{\otimes n}$,
where $|\psi^{+}\rangle\equiv\frac{1}{\sqrt{2}}(|00\rangle+|11\rangle)_{AB}$,
and transmits all the second qubits (say $B$) of the Bell pairs to
Bob, keeping the other half ($A$) with herself.
\item [{DLL2}] Bob randomly chooses a set of $\frac{n}{2}$ qubits from
the string received by him to form a verification string, on which
the BB84 subroutine to detect eavesdropping is applied. If sufficiently
few errors are found, they proceed to the next step; else, they return
to \textbf{DLL1}.
\item [{DLL3}] Alice randomly chooses half of the qubits in her possession
to form the verification string for the next round of communication,
and encodes her message in the remaining $\frac{n}{4}$ qubits. To
encode a 2-bit key message, Alice applies one of the four Pauli operations
$I,X,iY,Z$ on her qubits. Specifically, to encode $00,01,10$ and
$11$ she applies $I,X,iY$ and $Z$, respectively. After the encoding
operation, Alice sends all the qubits in her possession to Bob.
\item [{DLL4}] Alice discloses the coordinates of the verification qubits
after receiving authenticated acknowledgment of receipt of all the
qubits from Bob. Bob applies a BB84 subroutine to the verification
string and computes the error rate.
\item [{DLL5}] If the error rate is tolerably low, then Bob decodes the
encoded states via a Bell-state measurement on the remaining Bell
pairs.
\end{description}
DLL protocol described in this way helps us to illustrate the symmetry
among PP, CL and DLL protocols. This is a one-way
two-step QSDC protocol. DLL protocol looks similar to PP protocol
with dense coding (i.e., CL protocol). However, there is a fundamental
difference between a two-way protocol and a two-step one-way protocol
which uses the same resources and encoding operations. The difference
lies in the fact that in a two-way protocol home qubit always remains
at sender's port but in a one-way two-step protocol both the qubits
travel through the channel. At this specific point we observe a symmetry
between DLL protocol and GV protocol. Here the superposition is broken
into two pieces in such a way that the entire superposed (entangled)
state is never available in the transmission channel but only the
entire superposition (i.e., the superposed state or entangled state)
contains meaningful information. Visualization of this intrinsic symmetry
helps us to generalize DLL protocol to obtain an orthogonal version
of GV protocol.

\subsubsection{The modified DLL protocol (DLL$^{{\rm GV}}$)}

Based on the reasoning analogous to the one used for turning PP to
PP$^{{\rm GV}}$, we may propose the following GV-like version of
DLL, which may be called DLL$^{{\rm GV}}$ in accordance with the
recent work of Yadav, Srikanth and Pathak \cite{preeti-arxiv}. As
before, we retain the steps of DLL, replacing only steps \textbf{DLL1},
\textbf{DLL2} and \textbf{DLL4} as follows \cite{my book,preeti-arxiv}:
\begin{description}
\item [{DLL$^{{\rm GV}}$1}] Alice prepares the state $|\psi^{+}\rangle^{\otimes n}$.
She keeps half of the first qubits of the Bell pairs with herself.
On the remaining $\frac{3n}{2}$ qubits she applies a random permutation
operation $\Pi_{\frac{3n}{2}}$ and transmits them to Bob; $n$ of
the transmitted qubits are Bell pairs while the remaining $\frac{n}{2}$
are the entangled partners of the particles remaining with Alice.
\item [{DLL$^{{\rm GV}}$2}] After receiving Bob's authenticated acknowledgment,
Alice announces $\Pi_{n}\in\Pi_{\frac{3n}{2}}$, the coordinates of
the transmitted Bell pairs. Bob measures them in the Bell basis to
determine if they are each in the state $|\psi^{+}\rangle$. If the
error detected by Bob is within a tolerable limit, they continue to
the next step. Otherwise, they discard the protocol and restart from
\textbf{DLL$^{{\rm GV}}$1}.
\item [{DLL$^{{\rm GV}}$4}] Same as \textbf{PP$^{{\rm GV}}$4}, except
that the 'return trip' is replaced by Alice's second onward communication.
\end{description}
So two-way protocols of QSDC are now converted to one-way protocols.
But still we need two steps. This motivates us to ask: Do we always
need at least two steps for secure direct quantum communications?
Apparently it looks so because if we send both the qubits of an entangled
pair together then Eve may perform Bell measurement and find out the
message. Even if Eve is detected afterward it would not be of any
use because she has already obtained the message. However, using rearrangement
of particle order (PoP) we can restrict Eve from measuring in Bell
(special) basis and circumvent this problem. We have already used
PoP in implementing PP$^{{\rm GV}}$, CL$^{{\rm GV}}$ and DLL$^{{\rm GV}}.$
Using PoP a one-step one-way protocol of DSQC is already provided
by us in Ref. \cite{beyond-gv}. However, due to space restriction
we do not elaborate the one-step one-way orthogonal-state-based protocol
here.

We end-up this section by drawing your attention to the fact that
in all the existing protocols information splitting is done in such
a way that Eve does not get access to the special basis. Thus unavailability
of special basis leads to no-cloning and thus to secure quantum communication
and in the above described orthogonal-state-based protocol we have
primarily ensured unavailability of special basis by geographically
separating a quantum state into two pieces and avoiding Eve's simultaneous
access to both the pieces.

\section{Conclusions\label{sec:Conclusions}}

In the present work we have briefly reviewed the recent developments
on orthogonal-state-based protocols of secure quantum communication.
We have classified the recently proposed orthogonal-state-based protocols
into two sub-classes: GV-type and Counterfactual. GV-type protocols
are discussed with relatively more detail and it is explicitly shown
that by using a GV-type subroutine where Bell states are used as decoy
qubits we can convert any conjugate coding based protocol with orthogonal-state-based
encoding and decoding into GV-type completely orthogonal-state-based
protocol. Thus in principle, every task that can be done using conjugate
coding can also be done using orthogonal states alone. As examples,
we have explicitly shown here how PP and DLL protocols can be converted
to corresponding GV-type protocols. Further, since earlier proposals
of orthogonal-state-based protocols are experimentally implemented
recently, we may hope that ideas presented in this work and our more
detailed related works \cite{qka-chitra,beyond-gv,preeti-arxiv,dsqc-ent swap}
will be implemented soon and this type of protocols would draw much
more attention of cryptography community because of their fundamentally
different nature.

\textbf{Acknowledgment:} A .P. thanks Department of Science and Technology
(DST), India for support provided through the DST project No. SR/S2/LOP-0012/2010.
He also thanks K. Thapliyal for carefully reading the manuscript and
for helping in preparation of the figures. A. P. and R. S. thank N.
Alam, P. Yadav, A. Shenoy and S. Arvinda for their contribution on
the research works of the group that are reviewed in the present paper. Authors dedicate this work to Prof. Jozef Gruska on his 80th birth day.

\end{document}